\documentclass[aps,prd,twocolumn,nofootinbib]{revtex4-1}
\usepackage{graphicx}
\usepackage{amsmath}
\usepackage{graphicx}
\usepackage{amssymb}
\usepackage{color}
\usepackage{multirow}

\def\modify#1{#1} %use this to remove the highlight color

\def\be{\begin{equation}}
\def\ee{\end{equation}}
\def\bq{\begin{eqnarray}}
\def\eq{\end{eqnarray}}
\def\page{p_{\rm age}}

\begin{document}

\title{\modify{The $S_8$ Tension in Light of Updated Redshift-Space Distortion Data and PAge Approximation}}

\author{Lu Huang}
\affiliation{School of Physics and Astronomy, Sun Yat-sen University, 2 Daxue Road, Tangjia, Zhuhai, 519082, P.R.China}
\author{Zhiqi Huang$^*$}
\affiliation{School of Physics and Astronomy, Sun Yat-sen University, 2 Daxue Road, Tangjia, Zhuhai, 519082, P.R.China}
\email{huangzhq25@mail.sysu.edu.cn}
\author{Huan Zhou}
\affiliation{School of Physics and Astronomy, Sun Yat-sen University, 2 Daxue Road, Tangjia, Zhuhai, 519082, P.R.China}
\author{Zhuoyang Li}
\affiliation{School of Physics and Astronomy, Sun Yat-sen University, 2 Daxue Road, Tangjia, Zhuhai, 519082, P.R.China}

\begin{abstract}

  One of the most prominent challenges to the standard Lambda cold dark matter ($\Lambda$CDM) cosmology is the tension between the structure growth parameter $S_8$ constrained by the cosmic microwave background (CMB) data and the smaller one suggested by the cosmic shear data. Recent studies show that, for $\Lambda$CDM cosmology, redshift-space distortion (RSD) data also prefers a smaller $S_8$ that is $\sim 2$-$3\sigma$ lower than the CMB value, but the result is sensitive to the cosmological model. In the present work we update the RSD constraint on $S_8$ with the most up-to-date RSD data set where the correlation between data points is properly taken into account. To reduce the model dependence, we add in our Monte Carlo Markov Chain calculation the most up-to-date data sets of Type Ia supernovae (SN) and baryon acoustic oscillations (BAO), whose correlation with RSD is also taken into account, to constrain the background geometry. For $\Lambda$CDM cosmology we find $S_8= 0.812 \pm 0.026$, which is $\sim 2\sigma$ larger than previous studies, and hence is consistent with the CMB constraint. By replacing $\Lambda$CDM with the Parameterization based on cosmic Age (PAge), an almost model-independent description of the late universe, we find that the RSD + SN + BAO constraint on $S_8$ is insensitive to the cosmological model.

keywords:  observational cosmology, large-scale structure of the Universe, dark matter

  
\end{abstract}

\maketitle

\section{Introduction \label{sec:intro}}

The widely accepted explanation of cosmic accelerating expansion is that a dark energy component with negative pressure powers the late-time cosmic acceleration. The  $\Lambda$ cold dark matter ($\Lambda$CDM)  model, where the dark energy is interpreted as the cosmological constant $\Lambda$, has achieved great success in fitting a broad range of cosmological measurements in the last two decades~\citep{SN:1998fmf,Perlmutter1998MeasurementsO,Aghanim:2018eyx,eBOSS:2020yzd,Heymans:2020gsg,Asgari:2020wuj}. In recent few years, however, as the observational techniques continue to advance, the $\Lambda$CDM model is challenged by a few observational tensions, of which the two most prominent ones are the Hubble tension and the $S_8$ tension. The Hubble constant $H_0$ and the structure growth parameter $S_8$ measured by the Planck Satellite CMB experiment~\cite{Aghanim:2018eyx} are in $\sim 4\sigma$  tension with $H_0$ from the SH0ES distance ladder measurement~\cite{Riess:2020fzl} and in $\sim 3\sigma$ tension with $S_8$ from the cosmic shear data of the Kilo-Degree Survey~\cite{Asgari:2020wuj}, respectively. While many beyond-$\Lambda$CDM models are proposed as tentative explanations to the observed tensions, no specific model has been proven far better than $\Lambda$CDM~\cite{DiValentino:2021izs,EditorialCai2020, Guo2020, Liu:2019awo, Aghanim:2018eyx}.

The $3\sigma$ tension between CMB and cosmic shear in the inferred value of $S_8$ may arise from unaccounted baryonic physics~\cite{Lu:2021uvr}, other unknown systematics~\cite{Chintalapati:2021idc}, or a statistical fluke. It is important to have an $S_8$ probe that is independent of CMB and cosmic shear. Recent studies~\cite{Benisty:2020kdt,Nunes:2021ipq} suggest that the redshift-space distortion (RSD) data also prefer a small $S_8$ that is $\sim 2$-$3\sigma$ lower than the Planck result. However, the RSD constraint on $S_8$ is sensitive to the cosmological model. Ref.~\cite{Benisty:2020kdt} shows that a model-independent (Gaussian process) analysis yields a much weaker constraint on $S_8$ that is not very useful for resolving the $S_8$ tension.

The present work aims at updating the RSD constraint on $S_8$ with the most up-to-date RSD data sets. Because $S_8$ is degenerate with the background geometry parameters, we use in addition the data sets of Type Ia supernovae (SN)\modify{~\citep{Scolnic:2017caz}} and baryon acoustic oscillations (BAO)\modify{~\citep{Alam:2020sor}}  to determine the background geometry. In addition to the standard analysis for $\Lambda$CDM cosmology, we also perform our analysis for the Parameterization based on cosmic Age (PAge), which is an almost model-independent parameterization of the late universe~\cite{Huang:2020mub,Luo:2020ufj,Huang:2020evj,Cai:2021weh,MAPAge}. To compute the growth parameter in PAge, we assume that the clustering of the non-matter component is negligible in the late universe. \modify{This assumption in the present work excludes clustering dark energy models, which are typically very model-dependent and often studied in a model-by-model manner~\cite{Abramo07, Batista13, Batista17,Hassani19,Herrera19, Velten20, Creminelli20, Hassani21}.}

Unless otherwise specified, we work with a spatially flat Friedmann–Robertson–Walker background metric with scale factor $a=\frac{1}{1+z}$ and Hubble parameter $H = \frac{da/dt}{a}$, where $z$ is cosmological redshift and $t$ is the cosmological time. The structure growth parameter $S_8$ is defined as
\begin{equation}
  S_8\equiv \sigma_8\left(\frac{\Omega_m}{0.3}\right)^{0.5}, \label{eq:S8}
\end{equation}
where $\Omega_m$ is the matter density fraction at redshift zero. The root mean square of the matter density fluctuation $\sigma_8(z)$ is defined in a spherical top-hat window with comoving radius $8h^{-1}\mathrm{Mpc}$ at redshift $z$. Eq.~\eqref{eq:S8} reflects the main degeneracy direction of $\Omega_m$ and $\sigma_8$ parameters in cosmic shear surveys, where $\sigma_8$ without redshift argument implicitly refers to $\sigma_8(z=0)$.

This article is organized as follows. We briefly introduce PAge in Section~\ref{sec:page} and the updated data sets in Section~\ref{sec:data}. In Section~\ref{sec:results} we present the results and discuss their implications. Section~\ref{sec:conclusion} concludes.

\section{PAge Approximation \label{sec:page}}

PAge models the late-time cosmological evolution under two assumptions: i) the high-redshift universe is dominated by matter; ii) the dimensionless combination $Ht$ varies slowly and can be approximated as a quadratic function of $t$.  It follows from the two assumptions and general relativity that
\be
H = H_0 \left(1+\frac{2}{3}\left(1-\eta\frac{H_0 t}{\page}\right)\left(\frac{1}{H_0 t}-\frac{1}{\page}\right)\right), \label{eq:page}
\ee
where $\page=H_0t_0$ is the dimensionless age of the universe and $\eta < 1$ is a phenomenological parameter. Roughly speaking, $\eta$ characterizes the deviation from Einstein de-Sitter universe (flat CDM model)~\cite{MAPAge}.

By integrating Eq.~\eqref{eq:page}, we obtain the explicit expression of the scale factor
\begin{equation}
a(t) =\left ( \frac{H_0t}{\page} \right )^{\frac{2}{3}}e^ { \frac{\eta}{3}  \left ( \frac{H_0t}{\page}-1 \right )^{2}+
\left ( \page-\frac{2}{3} \right )\left ( \frac{H_0t}{\page}-1 \right )}. \label{ATrelation}
\end{equation}
Numeric inverse function of the right hand side of Eq.~\eqref{ATrelation} gives a mapping $t(a)$. The Hubble parameter at redshift $z$ is then obtained by substituting $t\left(\frac{1}{1+z}\right)$ into Eq.~\eqref{eq:page}. Integrating $\frac{dz}{H(z)}$ yields comoving angular diameter distance, which can be straightforwardly converted to the observable luminosity distance and angular diameter distance.

Further assuming that the clustering of non-matter component is negligible in the late universe, we can use the linear growth equation
\be
 \frac{d^{2}D }{dt^{2}}+2H\frac{dD }{dt}-\frac{3H_{0}^{2}}{2a^{3}}\Omega _{m}D =0 \label{eq:growthfactor}
 \ee
to evolve the linear growth factor $D$ of matter density fluctuations, at the price of introducing an extra parameter $\Omega_m$. The redshift-space distortion data measure the combination $f(z)\sigma _{8}(z) = f(z)D(z)\sigma_8(z=0)$, where the linear growth rate $f \equiv \frac{d\ln D}{d\ln a}$.

Refs.~\cite{Huang:2020mub,Luo:2020ufj,Huang:2020evj} have shown that many physically motivated or phenomenological models can be approximately mapped to PAge, with only $\sim$ sub-percent errors in the distance observables. Here we demonstrate the good accuracy in the observable $f\sigma_8$ with PAge approximation. A given model can be approximately mapped to $(\page, \eta)$ space by matching the cosmic age and the decelerating parameter $q_0 = -\frac{a \frac{d^2a}{dt^2}}{\left(\frac{da}{dt}\right)^2}$ at redshift zero. For a few typical examples, Table~\ref{table:comparison} shows the maximum relative error of $f\sigma_8$ and angular diameter distance $D_A$  in the redshift range $0<z<2.5$. 

\begin{table*}
  \caption{\label{table:comparison} Maximum relative errors of $D_A$ and $f\sigma_8$ in redshift range $0< z< 2.5$. }
  \centering
\begin{tabular}{cccccc}
\hline\hline
models &  parameters & $ \page$ & $\eta$ &$\max\left | \frac{\Delta D_A}{D_A} \right |\left(\% \right)$  &$\max\left | \frac{\Delta f\sigma_8}{f\sigma_8} \right |\left(\% \right)$ \\
\hline\hline
CDM & $\Omega_m=1$ & $\frac{2}{3}$ & 0& 0&0\\

flat $\Lambda$CDM & $\Omega_m=0.315$& 0.951& 0.359&0.416&0.874\\

flat $w$CDM & $\Omega_m=0.315,w=-1.2$ & 0.976& 0.619&0.579&1.07\\

flat $w_0$-$w_a$CDM & \ $\Omega_m=0.315,w_0=-1,w_a=0.3$\  & 0.941& 0.373&0.249&0.504\\

\hline\hline
\end{tabular}
\end{table*}

\begin{figure}
\centering
\includegraphics[width=0.47\textwidth]{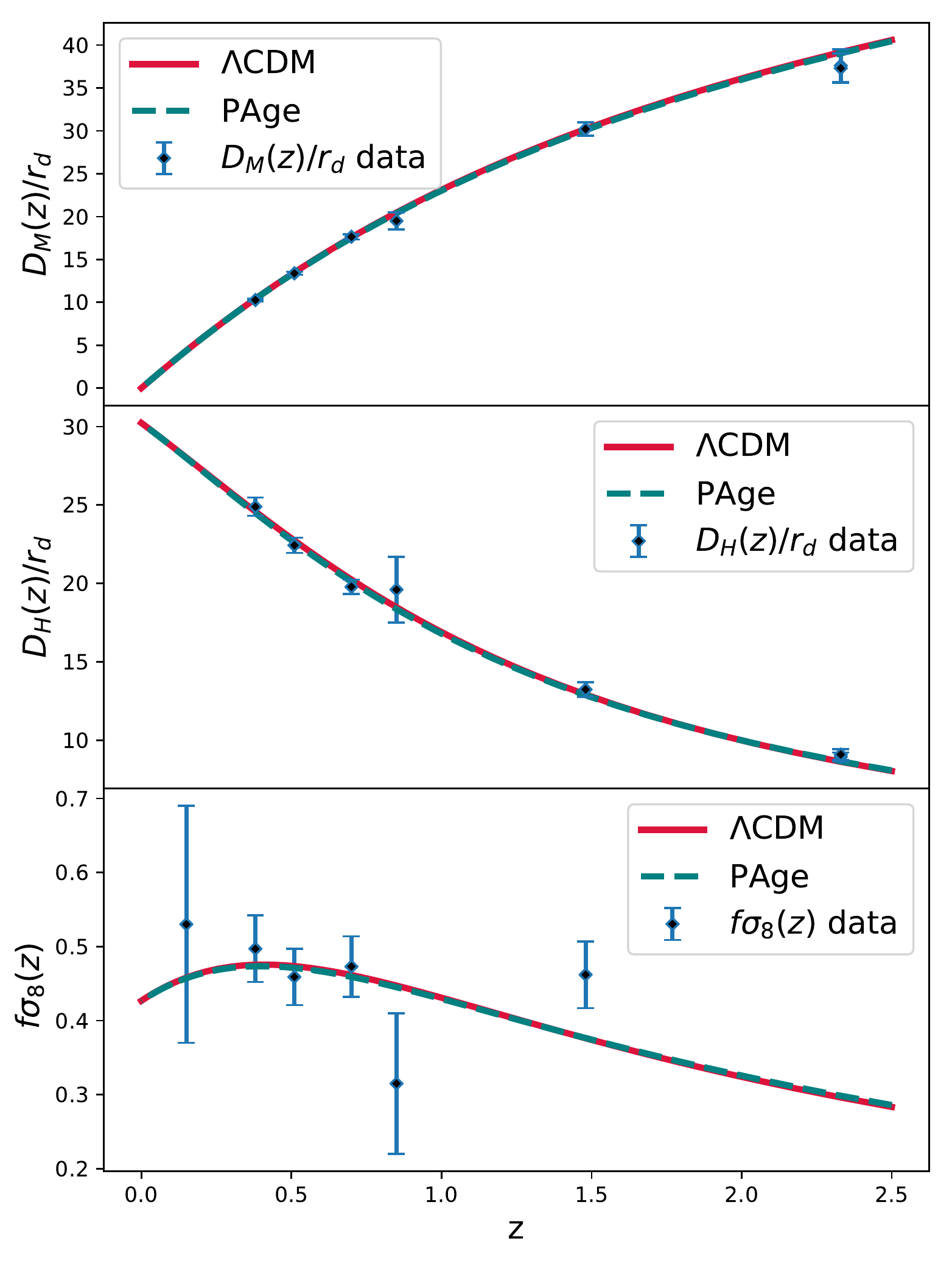}
\caption{The geometric measurements and growth rate measurements of SDSS-IV data. The red solid lines are predictions of $\Lambda$CDM model which assumes the best-fit results of Planck experiments~\citep{Aghanim:2018eyx} as fiducial values. The green dashed lines are their PAge approximations. Here $D_M$ is the comoving angular diameter distance, $D_H=\frac{c}{H}$, and $r_d$ is the comoving sound horizon at the end of the baryonic-drag epoch.}
\label{fig:model-data}
\end{figure}

In Figure~\ref{fig:model-data} we show some typical data points used in this article. The data uncertainties are much larger than the modeling errors of PAge approximation. Thus, for the presently achievable data precision, PAge can accurately represent many models in a very compact parameter space.

\section{Data and Methodology \label{sec:data}}

The growth rate data in the form of $f\sigma_8$ from RSD measurements have been widely used to study the $S_8$ tension~\citep{Nesseris:2017vor,Kazantzidis:2018rnb,Sagredo:2018rvc,Sagredo:2018ahx,Anagnostopoulos:2019miu,Skara:2019usd,Li:2019nux,Benisty:2020kdt,Nunes:2021ipq}. We carefully select the RSD measurements from different surveys to construct the $f\sigma_8$ data set, as listed in Table~\ref{table:data}. The data set contains the latest SDSS-IV and ALFALFA releases that are not included in the previous studies. We have also avoided using data points with unknown correlations that are possibly non-negligible due to redshift and sky-area overlap.
 
 \begin{table*}
\caption{\label{table:data}data }\centering
\begin{tabular}{llccccccll}
\hline\hline
Index &  Dataset& Redshift  & $f\sigma_8\left(z \right)$  & Fiducial Cosmology & Refs.\\
\hline\hline
1 & 2MTF & 0.001 & $0.505\pm0.085$ & $\Omega_{m} =0.312,\sigma_8=0.815$ & ~\citep{Howlett:2017asq} \\
2 & ALFALFA & 0.013 & $0.46\pm0.06$ & $\Omega_{m} =0.315,\sigma_8=0.8 $ &~\citep{Avila:2021dqv}\\
3 & 6dFGS+SnIa & 0.02 & $0.428\pm0.0465$ & $\Omega_{m} =0.3,\sigma_8=0.8 $&~\citep{Huterer:2016uyq}\\
4 & SNeIa+IRAS & 0.02 & $0.398\pm0.065$ & $\Omega_{m} =0.3,\sigma_8=0.814 $ & ~\citep{Turnbull:2011ty,Hudson:2012gt} \\
5 & 2MASS & 0.02 & $0.314\pm0.048$ & $\Omega_{m} =0.266,\sigma_8=0.65  $&~\citep{Hudson:2012gt,Davis:2010sw} \\
6 & 2dFGRS & 0.17 & $0.51\pm0.06$ & $\Omega_{m} =0.3,\sigma_8=0.9 $ & ~\citep{Song:2008qt}\\
7 & GAMA & 0.18 & $0.36\pm0.09$ & $\Omega_{m} =0.27,\sigma_8=0.8 $ & ~\citep{Blake:2013nif} \\
8 & GAMA & 0.38 & $0.44\pm0.06$ \\
9 & WiggleZ & 0.44 & $0.413\pm0.08$ & $\Omega_{m} =0.27,\sigma_8=0.8 $ & ~\citep{Blake:2012pj} \\
10 & WiggleZ & 0.60 & $0.39\pm0.063$ \\
11 & WiggleZ & 0.73 & $0.437\pm0.072$\\
12 & Vipers PDR-2& 0.60 & $0.55\pm0.12$& $\Omega_{m} =0.3,\sigma_8=0.823 $ &~\citep{Pezzotta:2016gbo} \\
13 & Vipers PDR-2& 0.86 & $0.40\pm0.11$ \\
14 & FastSound & 1.40 & $0.482\pm0.116$ & $\Omega_{m} =0.27,\sigma_8=0.82 $ &~\citep{Okumura:2015lvp} \\
15 & SDSS-MGS & 0.15 & $0.530\pm0.16$ & $\Omega_{m} =0.31,\sigma_8=0.83 $ &\modify{~\citep{Howlett:2014opa}}\\
16 & SDSS-BOSS-Galaxy & 0.38 & $0.497\pm0.045$ & $\Omega_{m} =0.31,\sigma_8=0.8$ & \modify{~\citep{BOSS:2016wmc}} \\
17 & SDSS-BOSS-Galaxy & 0.51 & $0.459\pm0.038$\\
18 & SDSS-eBOSS-LRG& 0.70  & $0.473\pm0.041$ & $\Omega_{m} =0.31,\sigma_8=0.8 $ & \modify{~\citep{Bautista:2020ahg,Gil-Marin:2020bct}}\\
19 & SDSS-eBOSS-ELG & 0.85  & $0.315\pm0.095$ & $\Omega_{m} =0.31,\sigma_8=0.8 $ & \modify{~\citep{deMattia:2020fkb,Tamone:2020qrl}}\\
20 & SDSS-eBOSS-Quasar & 1.48 & $0.462\pm0.045$ & $\Omega_{m} =0.31,\sigma_8=0.8 $ &\modify{~\citep{Hou:2020rse,Neveux:2020voa}} \\
\hline\hline
\end{tabular}
\end{table*}

In SDSS-IV survey, the systematic errors and  consensus estimates are incorporated directly into the covariance matrices. Therefore, for the \modify{MGS~\citep{Howlett:2014opa}, BOSS Galaxy~\citep{BOSS:2016wmc}, eBOSS LRG~\citep{Bautista:2020ahg,Gil-Marin:2020bct} and eBOSS Quasar~\citep{Hou:2020rse,Neveux:2020voa}} in BAO+RSD measurements, we take their covariance matrices into consideration to avoid missing correlation information. While for the \modify{eBOSS ELG~\citep{deMattia:2020fkb,Tamone:2020qrl}, Ly$\alpha$-Ly$\alpha$~\citep{duMasdesBourboux:2017mrl,Bautista:2017zgn} and Ly$\alpha$-Quasar~\citep{duMasdesBourboux:2020pck}} measurements, we directly introduce their publicly available likelihoods into our analyses. All of the likelihood information for the completed SDSS-IV are summarized on the public SDSS svn repository\footnote{ \url{https://svn.sdss.org/public/data/eboss/mcmc/trunk/
    likelihoods}}. The correlations of three WiggleZ data at different redshifts are considered as well.

Converting the redshift to distance by assuming a fiducial cosmology in RSD measurements leads to additional anisotropies known as Alcock-Paczynski (AP) effect. We multiply a correction factor to reduce the bias due to AP effect~\citep{Macaulay:2013swa}. The corrected $f\sigma_8$ for a model is 
\be
(f\sigma)^{\rm corrected}_{8} = \frac{H^{\mathrm{model}}\left ( z \right )D^{\mathrm{model}}_{A}\left ( z \right )}{H^{\rm fid}\left ( z \right )D_{A}^{\rm fid}\left ( z \right )}\times (f\sigma)^{\mathrm{model}}_{8}, \label{eq:APcorr}
\ee 
the superscript "fid" represents the fiducial flat $\Lambda$CDM cosmology assumed in RSD measurements. Eq.~\eqref{eq:APcorr} does not exactly eliminate the impact of AP effect. We will demonstrate that, however, AP correction has negligible impact on our results. Thus, the approximate correction in Eq.~\eqref{eq:APcorr} would suffice.

Finally, we construct our joint likelihood by multiplying the likelihoods of each survey in SDSS-IV, the WiggleZ survey,  the rest RSD surveys, and the Pantheon catalog, respectively. 

We use flat priors $hr_d\in[0,200]$, $\Omega_m \in [0,1]$ and $\sigma_8 \in [0.5,1.5]$ for $\Lambda$CDM model, and additionally $\page \in [0.8,1.2]$, $\eta \in [-1,1]$ for PAge. To reveal the impact of the AP correction, we also perform in parallel the analysis without the AP correction for a comparison.

\section{Results \label{sec:results}}

\begin{table*}
\caption{\label{table:constraint} The BAO + SN + RSD constraints on $\Lambda$CDM and PAge parameters}\centering
\begin{tabular}{llcccccccc}
\hline\hline
  &{\bf models} & $hr_{d}/\mathrm{Mpc}$ & $\page$ & $\eta$ &  $\Omega_m$   & $\sigma_8$ & $S_8$ & $ \modify{\chi^{2}_{min}}$ \\
\hline\hline
\multirow{2}{8em}{\bf without AP correction}& \bf $\Lambda$CDM & $100.8\pm1.1$&-&-& $0.289\pm0.011$& $0.823\pm0.025$&$0.807\pm0.027$ &\modify{1064.9}\\ 
& \bf PAge & $99.9\pm1.2$ & $0.969\pm0.012$ & $0.382\pm0.062$& $0.182^{+0.037}_{-0.058}$ & $1.07^{+0.12}_{-0.17}$&$0.808\pm0.026$ & \modify{1060.3}\\
\hline\hline
\multirow{2}{8em}{\bf with AP correction}& \bf $\Lambda$CDM &$100.7\pm1.1$ & - & -&$0.289\pm0.012$&$0.827\pm0.026$&$0.812\pm0.026$ & \modify{1065.6}\\
  &\bf PAge  &$100.1\pm1.2$&$0.970 \pm 0.012$& $0.391\pm0.063$&$0.180^{+0.034}_{-0.057}$&$1.07^{+0.12}_{-0.17}$&$0.807\pm0.026$& \modify{1060.8}\\
\hline\hline
\end{tabular}
\end{table*}

\begin{table*}
\caption{\label{table:repeat} The CC + SN + Gold-2017 constraints on $\Lambda$CDM parameters}\centering
\begin{tabular}{cccccc}
\hline\hline
$H_0$ & $\Omega_m$   & $\sigma_8$ & $S_8$ \\
\hline\hline
$69.7\pm1.9$ & $0.288\pm0.019$ & $0.774^{+0.031}_{-0.037}$ & $0.757\pm0.028$\\
\hline\hline
\end{tabular}
\end{table*}

We present the posterior mean and 68$\%$ confidence level inferences in Table~\ref{table:constraint}. Comparing the results with and without the AP correction, we conclude that the AP correction has little influence on the inferences of cosmological parameters. Thus, hereafter we only focus on the result with AP correction.

\modify{Because none of the BAO, SN, RSD data directly measures $H_0$, we can only obtain a combined constraint on $H_0r_d$, where $r_d$ is the comoving sound horizon at the end of the baryonic-drag epoch. Only when a Planck prior $r_d=147.21\pm 0.23\,\mathrm{Mpc}$~\cite{Aghanim:2018eyx} is used, can we obtain constraints on $H_0$ ($H_0 = 68.4\pm 0.8 \,\mathrm{km/s/Mpc}$ for $\Lambda$CDM and $H_0=68.0\pm 0.9\,\mathrm{km/s/Mpc}$ for PAge).}

Compared to the previous works~\citep{Nesseris:2017vor,Kazantzidis:2018rnb,Sagredo:2018rvc,Sagredo:2018ahx,Anagnostopoulos:2019miu,Skara:2019usd,Li:2019nux,Benisty:2020kdt,Nunes:2021ipq}, the present work prefers a higher $S_8$ value. The constraints $S_8=0.812\pm 0.026$ for $\Lambda$CDM and $S_8=0.807\pm 0.026$ for PAge are both consistent with CMB + $\Lambda$CDM measurement. The difference between previous work and ours may originate from the new RSD data and BAO data we have included in our analysis. To test this hypothesis, we utilize another catalog that contains cosmic chronometer (CC) data~\cite{Simon:2004tf,Stern:2009ep,Zhang:2012mp,Moresco:2012jh,Moresco:2015cya,Moresco:2016mzx,Ratsimbazafy:2017vga,Luo:2020ufj}, the Pantheon Type Ia supernova samples~\cite{Scolnic:2017caz} and the ``Gold-2017'' RSD compilation in Ref.~\cite{Nesseris:2017vor}. The inference results are listed in Table~\ref{table:repeat}. We find the previous ``Gold-2017'' RSD data do support a lower $S_8$ value that is consistent with previous works.

\begin{figure*}
\centering
\includegraphics[width=0.47\textwidth]{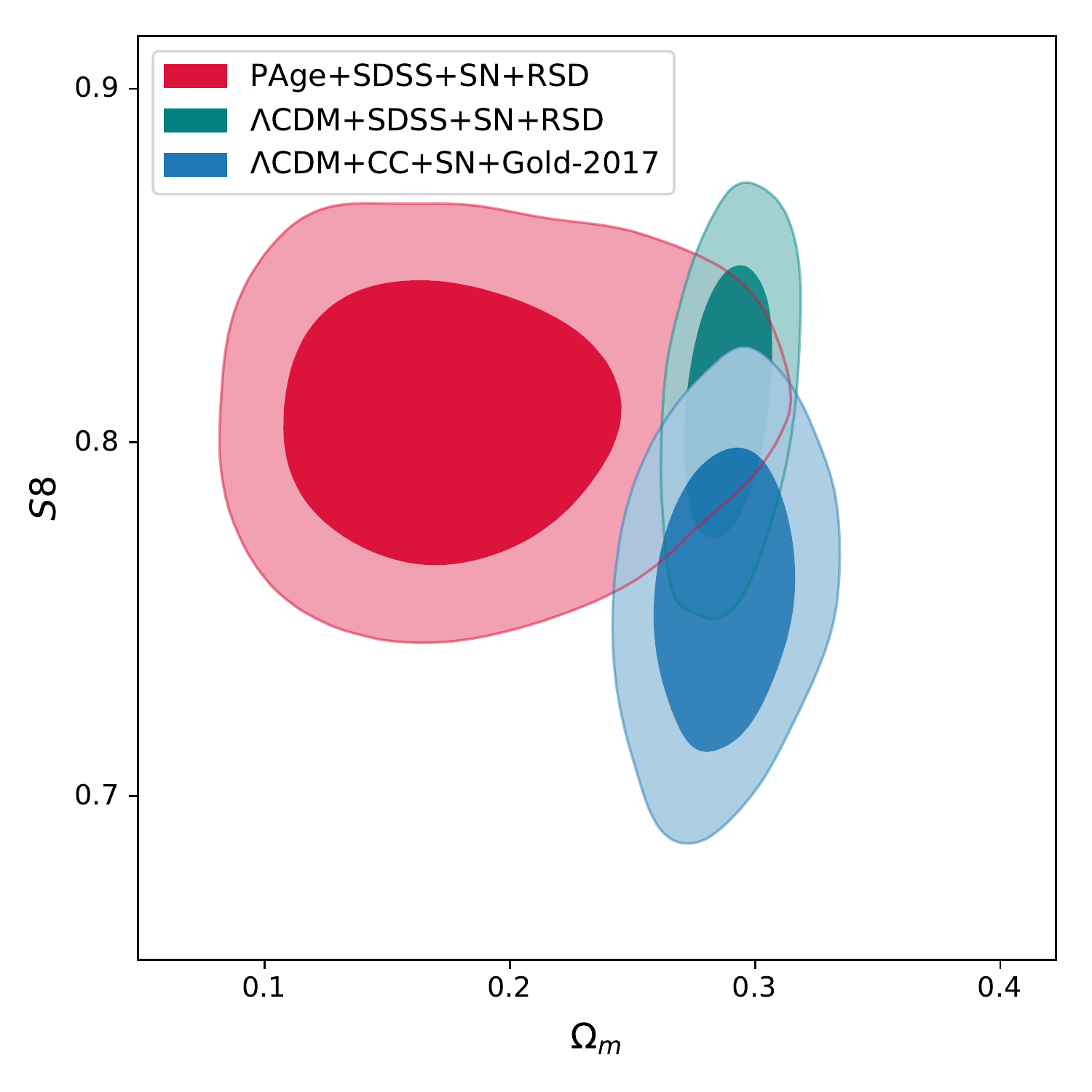}\includegraphics[width=0.47\textwidth]{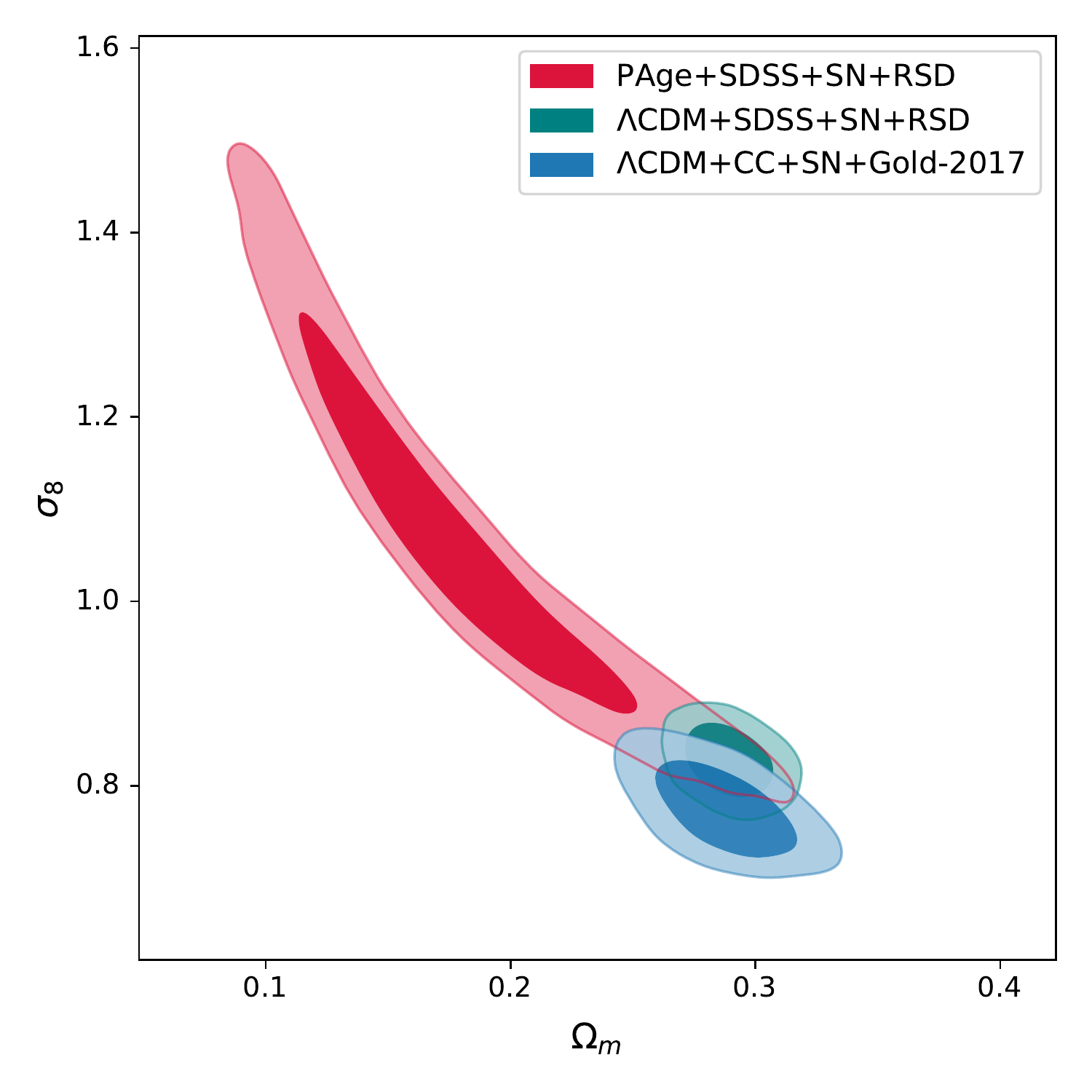}
\caption{Marginalized 68.3\% and 95.4\% confidence-level constraints. The SDSS data set include BAO + RSD from the latest SDSS-IV release, which are not included in the Gold-2017 RSD samples.}
\label{fig:contour}
\end{figure*}

Figure~\ref{fig:contour} shows the marginalized joint constraints on the $\Omega_m$-$S_8$ and $\Omega_m$-$\sigma_8$ planes. Note that in PAge $\Omega_m$ is only used to evolve the linear growth factor $D$. Thus, there is a strong degeneracy between $\sigma_8$ and $\Omega_m$ in the PAge case. In $\Lambda$CDM, the much tighter constraint on $\Omega_m$ from the measurement of background evolution breaks such degeneracy. \modify{The discrepancies of $\Omega_m$ and $\sigma_8$ parameters in different cosmologies, as shown in the right panel of Figure~\ref{fig:contour}, are $2.3\sigma$ and $1.6\sigma$, respectively. The seemingly large difference in minimal $\chi^2$, $5$ per degree of freedom, originates from a non-realistic parameter range ($\Omega_m\sim 0.18$ that is strongly disfavored by other cosmological observations~\cite{Aghanim:2018eyx}), and hence should not be regarded as a significant preference of PAge against $\Lambda$CDM.} The left panel of Figure~\ref{fig:contour} shows that, however, the uncertainty in $\Omega_m$ in PAge has no much impact on $S_8$ measurement, indicating that for RSD data the combination Eq.~\eqref{eq:S8} also roughly eliminates the degeneracy between $\Omega_m$ and $\sigma_8$.

\section{Conclusions and Discussion \label{sec:conclusion}}

RSD is an independent probe that is expected to help resolve the $S_8$ tension between CMB and cosmic shear measurements. However, the currently available RSD data are still very limited, and their constraint on $S_8$ is sensitive to the assumed background evolution model. In this work, we construct a clean and most up-to-date RSD catalog and use BAO + SN to eliminate the uncertainty of background evolution. The $S_8$ value inferred from our updated catalog is higher than many previous works. We have shown that it is mainly due to the inclusion of new RSD data and the usage of BAO + SN, which breaks the degeneracy between $S_8$ and the background evolution. \modify{Both the inference results of $S_8$ in $\Lambda$CDM and PAge agree well with the most updated result  $S_8=0.797^{+0.015}_{-0.013}$ of Dark Energy Survey (DES Y3)~\citep{DES:2021epj}. And they are found to be more consistent with CMB + $\Lambda$CDM measurement while the previous works~\citep{Nesseris:2017vor,Kazantzidis:2018rnb,Sagredo:2018rvc,Sagredo:2018ahx,Anagnostopoulos:2019miu,Skara:2019usd,Li:2019nux,Benisty:2020kdt,Nunes:2021ipq} tend to give a smaller $S_8$ that is more consistent with the cosmic shear data of the Kilo-Degree Survey~\cite{Asgari:2020wuj}.}

\modify{We have been working with a spatially flat cosmology. While including the spatial curvature parameter $\Omega_k$ into the analysis, we do not find significant impact on the $S_8$ constraint, which becomes $S_8=0.811\pm 0.027$ (almost unchanged) in the $\Lambda$CDM case and $S_8=0.800\pm 0.026$ (shifted by $0.3\sigma$) in the PAge case. This is because the background evolution is mostly data-driven (by SN+BAO) rather than theory-driven.}

In the PAge framework, the geometric information and structure-growth information are well split. The $\Omega_m$ parameter only affects the growth of structure via Eq.~\eqref{eq:growthfactor}. For the first time, we exclude the baryon-only ($\Omega_m=\Omega_b\approx 0.05$) and matter-only ($\Omega_m=1$) scenarios without assuming the Friedmann equations. This adds one more consistency of results from a broad variety of ways to observe aspects of the universe, which, in the view of Ref.~\cite{Peebles:2021gou}, makes the modern $\Lambda$CDM-like interpretation of the universe more robust. 

\modify{The growth of the structure beyond General Relativity has been intensively discussed in the literature~\citep{Kazantzidis:2018rnb,Anagnostopoulos:2019miu,Skara:2019usd,Nunes:2021ipq,DiValentino:2015bja,SolaPeracaula:2019zsl,SolaPeracaula:2020vpg,Yan:2019gbw}, some are found to be beneficial for relieving the Hubble tension and the $S_8$ tension~\cite{Yan:2019gbw}. In the present work, General Relativity is implicitly assumed in the evolution equation of the growth factor. Thus, alternative gravity theories where the Poisson equation of gravity is modified are beyond the scope fo this paper. We leave this as our future work.}

\section{Acknowledgements}

This work is supported by the National SKA Program of China No. 2020SKA0110402, National Natural Science Foundation of China (NSFC) under Grant No. 12073088, National key R\&D Program of China (Grant No. 2020YFC2201600), the science research grants from the China Manned Space Project with No. CMS-CSST-2021-B01, and Guangdong Major Project of Basic and Applied Basic Research (Grant No. 2019B030302001). 

%\bibliographystyle{apsrev4-1}
%\bibliography{ref}


\end{document}